# LLMorphism: When humans come to see themselves as language models

Valerio Capraro

University of Milano-Bicocca

valerio.capraro@unimib.it

**Abstract**

LLMorphism is the biased belief that human cognition works like a large language model. I argue that the rise of conversational LLMs may make this bias increasingly psychologically available. When artificial systems produce human-like language, people may draw a reverse inference: if LLMs can speak like humans, perhaps humans think like LLMs. This inference is biased because similarity at the level of linguistic output does not imply similarity in cognitive architecture. Yet, LLMorphism may spread through two mechanisms: analogical transfer, whereby features of LLMs are projected onto humans, and metaphorical availability, whereby LLM vocabulary becomes a culturally salient vocabulary for describing thought. I distinguish LLMorphism from mechanomorphism, anthropomorphism, computationalism, dehumanization, objectification, and predictive-processing theories of mind. I outline its implications for work, education, responsibility, healthcare, communication, creativity, and human dignity, while also discussing boundary conditions and forms of resistance. I conclude that the public debate may be missing half of the problem: the issue is not only whether we are attributing too much mind to machines, but also whether we are beginning to attribute too little mind to humans.

## The anthropomorphization of LLMs

For most of human history, the production of open-ended, context-sensitive, meaningful language was overwhelmingly associated with human speakers. It is therefore plausible that humans rely on a powerful heuristic: when an entity produces fluent and responsive language, we treat it as though there were a mind behind the words. The overapplication of this heuristic lies at the basis of the anthropomorphization of LLMs.

In general, anthropomorphism refers to the tendency to attribute human-like mental states, intentions, and capacities to non-human entities (Epley, Waytz, & Cacioppo, 2007). Anthropomorphism is not restricted to machines. It can be directed toward a wide range of non-human targets, including animals (de Waal, 1999), natural phenomena (Guthrie, 1995), supernatural agents (Barrett & Keil, 1996), and moving shapes (Heider & Simmel, 1944).

In the technological sphere, the anthropomorphization of machines did not begin with LLMs. Earlier work already showed that people readily apply social rules and expectations to computers, even when they know that such systems are artificial (Nass & Moon, 2000). However, LLMs may intensify anthropomorphism because they interact through language, a domain in which people are especially prone to adopt the intentional stance and infer communicative agency from apparently meaningful behavior (Dennett, 1989).

Conversational systems had already revealed this vulnerability. Weizenbaum's ELIZA, despite relying on a relatively simple pattern-matching architecture, elicited attributions of understanding, empathy, and involvement from users (Weizenbaum, 1966, 1976).

The contemporary case is more powerful because LLMs are far more fluent, flexible, and general. Shanahan (2024) explicitly warned that interaction with conversational LLMs can create the illusion of being in the presence of a thinking creature, even though such systems are not grounded in the world in the way humans are. Along these lines, empirical work suggests that users attribute mental states to LLMs, including capacities associated with experience (Chen et al., 2026; Cohn et al., 2024; Colombatto et al., 2025).

Thus, the first step is familiar: because LLMs produce human-like language, people may anthropomorphize them. However, LLMs also introduce a second and less discussed reverse inference.

## The LLMorphization of humans

As LLMs become widespread, a reverse inference becomes psychologically available: instead of merely asking whether LLMs are becoming human-like, people may begin to ask whether humans themselves are different from LLMs. This reverse inference may be adopted through at least two mutually reinforcing mechanisms: analogical transfer and metaphorical availability.

Analogical transfer involves mapping relational structure from one domain to another (Gentner, 1983). Similarity and analogy depend on structural alignment: when two systems appear similar, observers may align their relational structures and project inferences from one domain to the other (Gentner & Markman, 1997). In the present case, I suggest that the observable similarity between human and LLM language may make it tempting to reinterpret human linguistic competence through the lens of LLM-like text production. Because language is one of the main media through which human cognition becomes publicly visible, this analogy may extend beyond language itself: what begins as a comparison between two forms of linguistic output may become a broader reinterpretation of human cognition as LLM-like generation, prediction, recombination, and pattern completion.

Analogical transfer may be reinforced by the availability of cultural metaphors. Scientific and technological vocabularies often migrate into ordinary self-understanding, providing metaphors through which people conceptualize otherwise opaque psychological processes. Conceptual metaphor theory holds that people understand abstract domains partly by mapping them onto more concrete or culturally salient source domains (Lakoff & Johnson, 2008). Dominant technologies have repeatedly supplied models of the mind, from hydraulic and mechanical metaphors to computational and information-processing metaphors (Draaisma, 2000; Gigerenzer & Goldstein, 1996). Computers were not treated only as tools, but "objects to think with": artifacts through which people reflected on intelligence, aliveness, identity, and the nature of the self (Turkle, 2005). LLMs may provide the latest such vocabulary. Terms such as "prediction", "pattern completion", "generation", "prompting", "hallucination", "training data", and "next-token prediction" are not merely technical descriptors of artificial systems: they can become culturally available descriptors of human thought, creativity, memory, and explanation.

These two mechanisms may therefore lead to the reinterpretation of human cognition through the lens of LLMs. This is the *mechanomorphism* step. It does not consist simply in noticing that LLMs produce fluent language. It consists in using this fact to reinterpret human beings. The first move is justified: LLMs can produce language that resembles human language at the level of observable output. The second move is more questionable: because language is one of the main ways in which humans make thought socially visible, observers may treat similarity in language as evidence of similarity in cognition.

This inference, although tempting, is not logically valid, because similarity of linguistic outputs does not entail similarity of underlying cognitive processes (Loru et al., 2025; Quattrociocchi et al., 2025; Quattrociocchi et al., 2026). Human language is primarily a tool for communication developed by embodied organisms with needs, emotions, memories, social obligations, developmental histories, and vulnerability to consequences (Barsalou, 2008; Damasio, 2019; Gallagher, 2006; Fedorenko et al., 2024; Harnad, 1990; Seth, 2013; Tomasello, 2010). LLM output, by contrast, is generated by systems trained to model statistical regularities in language, rather than by embodied agents who use language from a situated perspective, with perception, action, affective stakes, communicative intentions, and responsibility for what they say (Bender et al., 2021; Shanahan, 2024; Zhao et al., 2023). Consequently, LLMs are remarkably strong at formal linguistic competence: they can generate grammatical, coherent, and locally context-appropriate text. However, they remain more limited with respect to functional linguistic competence, that is, the use and understanding of language in the world, which depends on extralinguistic capacities such as reasoning, world knowledge, situation modeling, and social cognition. For this reason, functional performance often requires forms of augmentation beyond next-word prediction, such as specialized fine-tuning or coupling with external modules (Mahowald et al., 2024).

The term *mechanomorphism* has a longer history than its current obscurity might suggest. Waters (1948) introduced it as “a new term for an old mode of thought”, defining it as the interpretation of human behaviour in terms of concepts and processes characteristic of machines. Subsequently, Caporael (1986) defined mechanomorphism as the attribution of machine characteristics to humans and explicitly contrasted it with anthropomorphism: anthropomorphism treats the machine as human-like, whereas mechanomorphism treats the human as machine-like. The concept has also appeared in human-animal studies (Karlsson, 2012).

While the term mechanomorphism is not new, it is also too broad for the current discussion. One may mechanomorphize humans by thinking of them as clocks, engines, robots, computers, algorithms, optimization systems, prediction machines. Instead, LLMs may give rise to a specific, and potentially more consequential form of mechanomorphism: LLMorphism, the biased belief that human cognition works like a large language model.

This specific bias may be particularly consequential because language is the medium through which cognition becomes socially visible: we display thought through explanations, arguments, stories, apologies, and justifications. Therefore, LLMorphism risks collapsing important distinctions between speaking and understanding, fluency and knowledge, generation and judgment. This makes it narrower than mechanomorphism, but potentially more consequential.

Importantly, LLMorphism does not consist in any comparison between humans and LLMs. Some comparisons are legitimate: humans also predict, compress information, generalize from experience, imitate others, and recombine learned patterns. The analogy becomes biased when these partial similarities are inflated into a general account of human cognition: when similarity in linguistic output is treated as sufficient evidence for similarity in cognitive architecture; when LLM-specific mechanisms are generalized to human cognition while ignoring embodiment, affect, agency, developmental history, social accountability, and non-linguistic thought.

Additionally, the claim is not that all people will explicitly endorse the theory that human cognition is equal to LLM cognition. It is weaker: as LLMs become widespread, some people may increasingly believe that human cognition is LLM-like.

### Difference between LLMorphism and adjacent constructs

We have already seen that LLMorphism is a specific form of mechanomorphism, and that mechanomorphism differs from anthropomorphism. In this section, I argue that LLMorphism is also distinct from other adjacent constructs.

LLMorphism is different from computationalism, the broader thesis that cognition is a form of information processing. Computationalism rests on the metaphor of the mind as a symbol-manipulating machine, operating over representations according to formal rules (Fodor, 1975; Newell & Simon, 1976). This view foregrounds logic, algorithmic problem-solving, and the execution of explicit procedures. It does not, in itself, imply any similarity to large language models. LLMorphism is not characterized by rule-based computation, but by the capacity of conversational LLMs to produce fluent, contextually appropriate, and human-sounding language. LLMorphism exploits the human heuristic that treats open-ended, coherent speech as a direct signal of a mind, which makes the conflation of output and process psychologically available.

LLMorphism also differs from dehumanization. Dehumanization involves the denial of humanness to individuals or groups. Haslam's model distinguishes animalistic dehumanization, which represents others as animal-like, from mechanistic dehumanization, which represents others as objects, automata, or machines (Haslam, 2006; Haslam & Loughnan, 2014). Mechanistic dehumanization is therefore close to mechanomorphism and LLMorphism. The key difference is that dehumanization typically concerns the denial of humanness to others, often with moral, interpersonal, or intergroup consequences (Opotow, 1990; Cuddy et al., 2007). LLMorphism, by contrast, need not involve hostility, contempt, moral exclusion, or the denial of human worth. Nor must it be directed only at others: one may LLMorphize humanity in general, or even oneself.

LLMorphism is also related to objectification, but again the overlap is partial. Objectification involves treating a person as an object, tool, or instrument rather than as a full subject (Nussbaum, 1995). LLMorphism may encourage objectification when people are seen as replaceable mechanisms or output-generating systems. However, LLMorphism does not necessarily involve using another person instrumentally. Its primary content is representational: it concerns how humans are conceptualized, not necessarily how they are exploited.

LLMorphism is also distinct from predictive processing and related Bayesian theories of cognition. Predictive processing holds that the brain continuously generates predictions about sensory input and updates internal models in light of prediction error (Clark, 2013; Friston, 2010; Hohwy, 2013). But predictive processing does not imply that humans are LLM-like, nor that human understanding is merely text generation. Indeed, many predictive-processing accounts are deeply embodied and action-oriented (Allen & Friston, 2018; Clark, 2015; Pezzulo et al., 2024).

**Potential impacts of LLMorphism on society**

The consequences of LLMorphism may extend beyond how people talk about AI and themselves. The following potential impacts should be understood as hypotheses rather than established facts. Since LLMorphism is introduced here as a new construct, there is not yet direct empirical evidence on its prevalence or effects. Some of the risks outlined below may turn out to be negligible; others may prove more consequential than currently anticipated. The aim of this section is therefore not to provide a definitive account, but to identify plausible pathways through which LLMorphism could impact social life and to open a discussion on how its negative consequences might be limited.

One possible pathway is a replaceability mechanism. Organizations often evaluate people through outputs: documents, reports, metrics, code, and productivity traces. If humans are seen as LLM-like output generators, automation may appear not only economically efficient, but conceptually justified. This risk is not new and and connects with broader debates on automation and labour substitution, which show that technologies can reorganize work around tasks that machines can perform or measure (Acemoglu & Restrepo, 2019; Autor, 2015; Parasuraman, & Riley, 1997), and with algorithmic management, where workers are governed through quantifiable productivity traces (Kellogg et al., 2020). Yet, LLMorphism may make humans appear even less grounded, less agentic, and more replaceable.

A second pathway is a fluency mechanism. Many forms of expertise become socially visible through language: lectures, diagnoses, legal arguments, reviews, policy briefs, and professional advice. If LLMs approximate these outputs, institutions may mistake expertise itself for sophisticated text production. This confuses the expression of expertise with the processes that make it reliable. Expert judgment depends on tacit knowledge, situated interpretation, uncertainty management, disciplinary norms, and accountability within a practice (Polanyi, 1966; Collins & Evans, 2019). In education, the parallel risk is that learning becomes equated with fluent answers. If fluent answers are treated as evidence of understanding, learning may become equated with the production of well-formed text, even though linguistic form is not equivalent to grounded meaning or functional competence (Bender & Koller, 2020). LLMorphism may therefore make fluency appear sufficient for understanding and, in doing so, devalue expertise and weaken educational norms.

A third pathway is an agency-thinning mechanism. Human beings are ordinarily treated as agents because they can respond to reasons, recognize obligations, anticipate consequences, and be called to account (Fischer & Ravizza, 1998; Scanlon, 1998). If human action is redescribed as generated output from prior inputs, agency may appear thinner. This does not mean that mechanistic explanations are incompatible with responsibility: human behaviour partly depends on biology, development, context, and constraint. The risk is rather that explanation becomes displacement. Describing action as pattern completion may replace questions about reasons, commitments, negligence, intention, and accountability with a flatter vocabulary of input, output, and generation (Brandom, 1994; Duff, 2007). In legal and moral life, this could erode the interpersonal foundations of blame, apology, repair, trust, and responsibility, the practices through which people address one another as accountable participants rather than merely as systems to be predicted, managed, or modified (McKenna, 2012; Walker, 2006).

A fourth pathway is a disembodiment mechanism. If cognition is increasingly understood through the model of language generation, verbal output may be over-weighted relative to embodied, affective, and situational cues. This risk is especially visible in healthcare. What patients say is crucial, but clinicians also rely on how patients appear. Research on clinical communication shows that nonverbal behaviour is central to physician–patient interaction, including the expression of emotion, empathy, distress, and relational understanding (Roter et al., 2006; Henry et al., 2012). More broadly, patient-centred medicine treats illness not merely as verbally reportable information, but as an embodied and socially situated experience (Engel, 1977; Kleinman, 1988; Carel, 2016). LLMorphism may therefore encourage a more linguistically biased medicine, in which patients are treated primarily as producers of clinically relevant text rather than as embodied persons whose condition must also be perceived, interpreted, and cared for. This could produce errors in both directions: fluent patients may be underestimated despite visible deterioration, while anxious, cognitively impaired, or non-native-speaking patients may be misread when their verbal reports are fragmented, ambiguous, or difficult to interpret (Flores, 2005; Karliner et al., 2007). The risk may be particularly acute in mental health, where suffering can be difficult to articulate and where coherent self-description does not always track clinical severity; behavioral and nonverbal signs such as psychomotor retardation, agitation, facial expression, vocal dynamics, and posture can provide clinically relevant information beyond verbal report (Dibeklioğlu et al., 2015).

A fifth pathway is epistemic. If humans are increasingly interpreted through the model of LLMs, cognition itself may come to be understood as fluent generation rather than grounded inquiry. This may reinforce epistemia: a condition in which linguistic plausibility substitutes for epistemic evaluation (Loru et al., 2025; Quattrociocchi et al., 2025). LLMorphism blurs the distinction between saying and knowing. If human reasoning is "just" pattern completion, then justification, evidence, and understanding may lose normative force. The result is not only that people may overtrust AI-generated text, but that human thought itself may be redescribed in terms that make fluency appear sufficient for knowledge. In this sense, LLMorphism may contribute to a broader epistemic shift: from evaluating whether claims are grounded, justified, and accountable, to evaluating whether they are coherent, fluent, and plausible.

**Boundary conditions**

The preceding sections have focused on mechanisms that may spread LLMorphism and on potential societal consequences. However, several countervailing forces, operating at cognitive, dispositional, and socio-contextual levels, are likely to moderate these impacts.

Analogical transfer, according to structure-mapping theory, requires that the source (LLM) and target (human cognition) be aligned on relevant relational predicates (Gentner & Markman, 1997). In the present case, LLMorphism should be facilitated when LLMs and human cognition are aligned around fluent language production. Conversely, it should be weakened when people attend to salient disanalogies. Interventions that make these disanalogies cognitively available may therefore reduce the tendency to interpret human cognition through an LLM-like frame (Wood et al., 2025).

Metaphorical availability may be constrained when alternative source domains compete for the description of human thought. Conceptual metaphor theory holds that abstract concepts are often understood through multiple metaphors, whose entailments can highlight different aspects of the same target domain and sometimes pull interpretation in different directions (Lakoff and Johnson, 2008). The cultural dominance of the LLM metaphor is not guaranteed: it can be challenged by reviving older metaphors or by introducing new ones. For example, individuals who strongly endorse essentialist beliefs about human uniqueness (Haslam, Bastian & Bissett, 2005) or who hold religious or humanistic worldviews that posit a non-material soul (Bloom, 2012) are likely to reject the LLM metaphor outright, substituting it with an incompatible framework.

Experiential and socio-contextual moderators may also be important. Professional roles involving direct caregiving, such as nursing, therapy, or early childhood education, repeatedly confront practitioners with aspects of human life that resist LLM-style description. Nursing involves emotional and relational labour that cannot be captured by textual output alone (Feng et al., 2024). Psychotherapy depends not only on verbal exchange, but also on embodied presence, nonverbal communication, and therapeutic alliance (Del Giacco et al., 2020; García et al., 2022). Early childhood education is organized around relational pedagogy, attachment, affect regulation, and development (Cliffe & Solvanson, 2023). These experiences may make human–LLM disanalogies more cognitively salient and thereby reduce the appeal of LLMorphism. Training in the humanities or qualitative social sciences may operate similarly, by emphasizing narrative, interpretation, lived experience, and first-person perspective (Charon, 2001; Neubauer et al., 2019).

**Discussion**

This paper introduced LLMorphism: the biased belief that human cognition works like a large language model. I argued that conversational LLMs may make this bias increasingly psychologically available by reversing a familiar heuristic: fluent, meaningful language has traditionally been treated as evidence of a human mind. Once LLMs produce fluent language, the inference may be reversed: if LLMs can speak like humans, perhaps humans think like LLMs. This inference is biased because similarity in linguistic output is treated as evidence of similarity in cognitive architecture. I suggested that LLMorphism may spread through two mutually reinforcing mechanisms: analogical transfer, whereby features of LLMs are projected onto humans, and metaphorical availability, whereby LLM vocabulary becomes a culturally salient vocabulary for describing thought. I also distinguished LLMorphism from adjacent constructs, including anthropomorphism, mechanomorphism, computationalism, dehumanization, objectification, and predictive theories of mind. I outlined several pathways through which LLMorphism may reshape social judgments: by making humans appear more replaceable, making fluency seem sufficient for expertise, thinning agency and responsibility, reducing attention to embodied and affective cues, and weakening the distinction between plausible generation and grounded understanding. Finally, I described potential boundary conditions and moderating factors.

LLMorphism is introduced here as a new construct, opening the way to several potential questions. The first task for future work is construct measurement. Future research should develop a psychometric scale capturing its potentially multidimensional structure. LLMorphism may not be a single belief, but a cluster of related intuitions. One dimension may concern language generation: the belief that human thought and speech are forms of next-token prediction. This dimension may descend from a reductionist interpretation of predictive and probabilistic models of human language processing, which show that humans anticipate upcoming linguistic input and that next-word prediction can partially model neural responses to language (Schrimpf et al., 2021; Goldstein et al., 2022). A second dimension may concern learning and creativity: the belief that humans learn primarily by extracting statistical regularities and that creativity is essentially recombination of existing material. This dimension may draw superficial support from broader combinatorial theories of creativity (Boden, 2004; Simonton, 2010). A third dimension may concern self-understanding and confabulation: the belief that introspection, explanation, and inner speech are post-hoc narratives. This belief may come from a reductionist extrapolation of research showing that people often have limited introspective access to the processes underlying their judgments (Nisbett & Wilson, 1977; Johansson et al., 2005; Wilson, 2002). A fourth dimension may concern truth-seeking: the belief that human reasoning is not primarily oriented toward truth, but toward producing plausible outputs. This belief may descend from a superficial interpretation of evidence showing that reasoning is often shaped by argumentative goals and motivated reasoning (Kunda, 1990; Mercier & Sperber, 2011). Other dimensions may emerge empirically. Developing and validating such a scale would be essential for determining whether LLMorphism is best understood as a unitary or multidimensional construct, or a family of partially overlapping beliefs.

Future work should also examine individual differences in susceptibility to LLMorphism. One possibility is that LLMorphism is stronger among people with greater exposure to LLMs. However, the opposite prediction is also plausible: greater technical understanding may reduce LLMorphism by making users more aware of the differences between human cognition and model architecture. Other potential difference may be searched along socio-economic lines that are known to influence LLM adoption, including gender (Carvajal et al., 2025), education (Yoon & Kim, 2025), occupation (Humlun & Vesterlund, 2025), among others (Capraro et al., 2024).

Finally, future work should test the social consequences of LLMorphism. The framework predicts that stronger LLMorphic beliefs will increase perceived replaceability of human workers, reduce perceived distinctiveness of human expertise, weaken attributions of agency and moral responsibility, increase reliance on verbal fluency as a proxy for understanding, and reduce attention to embodied, affective, and contextual cues in domains such as education, healthcare, law, and work. Experimental studies could manipulate exposure to LLMs and then measure whether participants become more likely to describe human cognition as it works like a language model.

Clearly, these are hypotheses: some may turn out to be false, others may need refinement, and still others may be discovered. The broader point, however, is that public debate on AI has focused mainly on anthropomorphism: whether we are giving too much mind to machines. LLMorphism suggests that this is only half of the problem. The other half is whether we are beginning to take too much mind, agency, and grounding away from humans.